\documentstyle[12pt]{article}

\begin{document}

\begin{titlepage}
\begin{flushright}
{\large \bf UCL-IPT-97-14 \\
\bf CINVESTAV-FIS/97-11}
\end{flushright}
\vskip 2cm
\begin{center}

{\Large \bf Tests for the asymptotic behaviour of the $\gamma^*
\rightarrow \gamma \pi^0$ form factor}
 \vskip 2cm

{\large J.-M. G\'erard$^{a}$ and G. L\'opez Castro$^b$} \\

$^a$ {\it Institut de Physique Th\'eorique, Universit\'e catholique de
Louvain,}\\ {\it  B-1348 Louvain-la-Neuve, Belgium} \\

$^b$ {\it Departamento de F\'\i sica, Centro de Investigaci\'on y de  
Estudios} \\ {it Avanzados del IPN, Apdo. Postal 14-740, 07000 M\'exico, 
D.F., M\'exico}

\end{center}

\vskip 2cm

\begin{abstract}
The $\gamma^* \rightarrow \omega \pi^0$ transition measured at different
photon virtualities already provides us with a clean test for the
behaviour of the $\pi^0 \gamma^* \gamma$ off-shell axial anomaly at
large
time-like squared momenta. It also allows reliable predictions for the
branching ratio of heavy quarkonium decays into $\omega \pi^0$. 
\end{abstract}

\


\end{titlepage}%

\medskip

\

Extrapolations of the anomalous $\pi^0 \gamma\gamma$ vertex to virtual
$\gamma^*$
photons are mandatory to study several physical processes.
In a previous letter \cite{jm-th}, the Bjorken-Johnson-Low theorem
\cite{bjl} was applied to derive the behaviour of the matrix element for
the $\gamma^* \rightarrow \gamma \pi^0$ transition when the incoming
$\gamma^*$ photon is very energetic. In the present note we would like to
emphasize
that the related $\gamma^* \rightarrow \omega\pi^0$ form factor has
already been measured at sufficiently large squared-momenta to test the
predicted asymptotic behaviour of the $\pi^0\gamma^*\gamma$ off-shell
anomaly in the time-like region.

\

{\bf 1. The $e^+e^- \rightarrow \omega\pi^0$ cross section}

The DM2 Collaboration \cite{dm2} has observed the $e^+e^- \rightarrow
\omega\pi^0$ exclusive channel up to $\sqrt{s} = 2.4 $ GeV. The associated
cross section is given by
\begin{equation}
\sigma (e^+e^- \rightarrow \omega\pi^0)\ = \ \frac{\pi \alpha^2}{6}
|F_{\omega\pi^0}(s)|^2 \Phi ^3(s)\ ,
\end{equation}
with $\sqrt{s}$ the center of mass energy and
\begin{equation}
\Phi (s) \equiv \left\{ 
\left[ 1-\frac{(m_{\omega} +m_{\pi^0})^2}{s} \right] 
\left[ 1-\frac{(m_{\omega} -m_{\pi^0})^2}{s} \right] \right\}^{1/2} \ ,
\end{equation}
the standard phase space factor. In Eq.(1), the $\gamma^* \rightarrow
\omega \pi^0$ form factor $F_{\omega \pi^0}(s)$ is defined as follows
\begin{equation}
\langle \omega(\varepsilon,\ q) \pi^0(q') | ej^{em}_{\mu}(0) |0 \rangle
\equiv ie F_{\omega \pi^0} (s) \epsilon_{\mu \nu \alpha \beta} q^{\nu}
q'^{\alpha} \varepsilon^{\beta}\ .
\end{equation}
Its momentum dependence can be extracted from the published DM2 data using
the
relation 
\begin{equation}
| F_{\omega \pi^0} (s) | = \left( \frac{6}{\pi \alpha^2} \right)^{1/2}
\Phi^{-3/2}(s) \left[ \sigma^{DM2} (e^+e^- \rightarrow \omega \pi^0)
\right]^{1/2}\ .
\end{equation}
  The resulting $\gamma^* \rightarrow \omega \pi^0$ form factor is given
in Fig.1 for 1.5 GeV $\leq \sqrt{s} \leq$ 2.4 GeV. At lower momenta,
the $\rho$ meson contribution dominates the cross section.
   This general property of vector dominance is nicely confirmed by the
complementary data \cite{aleph} on $\tau \rightarrow \nu_{\tau} \omega
\pi^-$ decays.

  From the DM2 data alone, we conclude that the $\gamma^* \rightarrow
\omega
\pi^0$ form factor decreases rather regularly with increasing $\sqrt{s}$.
But a definite conclusion would require more data above 2.4 GeV.
Fortunately, the
$J/\psi$ resonance allows us to extract one additional value of $F_{\omega
\pi^0}(s)$ at a higher value of $\sqrt{s}$.

\

{\bf 2. The $J/\psi \rightarrow \omega \pi^0$ decay width.}

\

The width of the quarkonium decay $V_Q \rightarrow \omega \pi^0$ is given
by
\begin{equation}
\Gamma (V_Q \rightarrow \omega \pi^0) = \frac{m_{V_Q}^3}{96\pi} |g_{V_Q
\omega \pi} |^2 \Phi^3 (m_{V_Q}^2) 
\end{equation}
in the rest frame of $V_Q$. For heavy ($ Q=c,b, \cdots$) quarkonia, this
isospin-violating decay is dominated by the single (iso-triplet) photon
exchange. The three-gluon exchange contributions followed by $\rho-\omega$
or $\eta-\pi$ mixing transitions are indeed negligible for this specific
channel. Consequently, the effective $g_{V_Q \omega \pi}$ coupling can be
expressed in terms of the $\gamma^* \rightarrow \omega \pi^0$ form
factor:
\begin{equation}
|g_{V_Q\omega \pi} | = \frac{4\pi \alpha}{f_{V_Q}} |F_{\omega\pi^0}
(m^2_{V_Q}) |
\end{equation}
with $f_V$, the vector decay constant, defined by
\begin{equation}
\langle 0| ej^{em}_{\mu} (0) | V(\varepsilon) \rangle \equiv
e\frac{m^2_V}{f_V} \varepsilon_{\mu}
\end{equation}
and extracted from the $V \rightarrow e^+e^-$ leptonic decay width:
\begin{equation}
\Gamma (V \rightarrow e^+e^-) = \frac{4\pi}{3} \left( \frac{\alpha}{f_V}
\right )^2 m_V\ .
\end{equation}
  
From the experimental rate \cite{pdg} for $J/\psi \rightarrow \omega
\pi^0$
we obtain
\begin{equation}
|F_{\omega \pi^0}(m^2_{J/\psi}) | = (8.45 \pm 0.62)\cdot 10^{-2}\ {\rm
GeV}^{-1}\ .
\end{equation}

This additional value for the $\gamma^* \rightarrow \omega \pi^0$ form
factor is included in Fig.1. It strongly suggests a rather precocious
asymptotic behaviour. A straightforward use of vector meson dominance
provides therefore a nice test for the behaviour of the $\pi^0 \gamma^*
\gamma$ off-shell axial anomaly at large time-like squared momenta which
has been derived \cite{jm-th} from the Bjorken-Johnson-Low (BJL) theorem
\cite{bjl}.

\

{\bf 3. The $\gamma^* \rightarrow \gamma \pi^0$ form factor.}

\

The BJL theorem applied to the $\gamma^* (q_1) \rightarrow
\gamma(q_2) \pi^0$ matrix element implies \cite{jm-th}
 \begin{equation}
{\cal M} (q_1^0 \rightarrow \infty) = e^2 \frac{\sqrt{2} f_{\pi}}{3
q_1^0}\ \epsilon_{\mu \nu 0 \sigma} (q_2 -q_1)^{\sigma}
\varepsilon^{\mu}_1
\varepsilon^{\nu}_2 
 \end{equation}
with $f_{\pi} = 131$ MeV, the pseudoscalar decay constant of the pion. For 
the processes considered in this paper, 
$\vec{q_1}=0$ because we work in the center of mass frame of the
$\omega\pi^0$ system. If we
multiply and divide the right-hand side of Eq.(10) by $q_1^0$ with
$(q_1^0)^2 = s$, we obtain
then the following asymptotic behaviour 
\begin{equation}
{\cal M} (s \rightarrow \infty) = e^2 \frac{\sqrt{2} f_{\pi}}{3s}\
\epsilon_{\mu\nu\rho\sigma} q_1^{\rho} q_2^{\sigma} \varepsilon_1^{\mu}
\varepsilon_2^{\nu}
\end{equation}
namely,
\begin{equation}
F_{\gamma \pi^0} (s\rightarrow \infty) = \frac{\sqrt{2} f_{\pi}}{3 s} .
\end{equation}
We notice that our normalization is such that
\begin{equation}
F_{\gamma \pi^0} (0) = -\frac{\sqrt{2}}{4\pi^2}\, \frac{1}{f_{\pi}} 
\end{equation}
is the form factor associated with the $\pi^0 \rightarrow \gamma \gamma$
triangle anomaly \cite{abj}.

  The $\omega-$vector dominance applied to the outgoing (iso-singlet)
real photon implies therefore the following asymptotic behaviour for the
$\gamma^*(I=1) \rightarrow \omega \pi^0$ form factor:
\begin{equation}
F_{\omega \pi^0} (s\rightarrow \infty) = \frac{f_{\omega} f_{\pi}}{3
\sqrt{2} s}
\end{equation}
with $f_{\omega} = 17.05 \pm 0.28$, the vector decay constant defined by
Eqs.(7) and (8). As illustrated in Fig.1, the asymptotic regime for the
$\gamma^* \rightarrow \omega \pi^0$ isotriplet form factor derived from
the BJL theorem (supplemented with vector dominance) is already almost
saturated at the $J/\psi$ mass scale. A crucial confirmation of this
remarkable property would require the observation of heavy quarkonium
decays into $\omega \pi^0$.

\

{\bf 4. The $\Upsilon(1S) \rightarrow \omega \pi^0$ branching ratio.}

\

The vector decay constant $f_{\Upsilon} = 40.0 \pm 0.8$ is estimated on
the basis of Eq.(8). It is then straightforward to predict
\cite{correction} the $\Upsilon \rightarrow \omega \pi^0$ branching
ratio from Eqs.(5), (6) and (14):
\begin{equation}
BR(\Upsilon \rightarrow \omega \pi^0) = (0.95 \pm 0.06) \cdot
10^{-5}\ .
\end{equation}

Needless to say that a measured branching ratio around $10^{-5}$ would
definitively support the behaviour given in Eq.(12) for the $\pi^0
\gamma^*\gamma$ off-shell
axial anomaly at large time-like momenta. 

   In conclusion, the asymptotic behaviour of the $\pi^0 \gamma^* \gamma$
off-shell axial anomaly at large time-like squared momenta derived in 
ref.\cite{jm-th} has been successfully tested through the study of the related
$\pi^0 \gamma^* \omega$ form factor measured at rather high energies.
Further
tests would require either data above 2.4 GeV for the $e^+ e^- \rightarrow
\omega \pi^0$ process or a measurement of the $\Upsilon \rightarrow \omega
\pi^0$ branching ratio at the $10^{-5}$ level.

 The asymptotic behaviour of the $\gamma \pi$ transition form factor given
in Eq.(12) is a factor 1/3 smaller than the one obtained \cite{bl1} from
the QCD infinite momentum frame approach for the description of the pion
wave function. Consequently, the data shown in Fig.1 do not support the
Brodsky-Lepage limit for {\it time-like} momentum transfers up to 9
GeV$^2$.

 On the other hand, the $\gamma \pi$ transition form factor measured by
the CELLO \cite{cello} and CLEO \cite{cleo} Collaborations for {\it
space-like} momentum transfers up to 8 GeV$^2$ is close to the
Walsh-Zerwas limit \cite{wz} and reproduces the Brodsky-Lepage
interpolating formula \cite{bl2,new}. However, we stress again that the
BJL theorem alone does not allow us to derive the asymptotic behaviour for
large space-like momenta.

 \

\newpage

\medskip

\begin{center}
FIGURE CAPTION
\end{center}
Energy dependence of the $\gamma^* \rightarrow \omega \pi^0$
form factor ($E=\sqrt{s}$): comparison of experimental data (points) with 
the prediction (solid curve) based on the BJL theorem and given in
Eq.(14).


\medskip

\newpage 

\begin{thebibliography}{99}

\bibitem{jm-th} J.-M. G\'erard and T. Lahna, Phys. Lett. {\bf B356} (1995)
381.

\bibitem{bjl} J. D. Bjorken, Phys. Rev. {\bf 148} (1966) 1467; K. Johnson
and F. E. Low, Suppl. Prog. Theor. Phys. {\bf 37} (1966) 74; see also
S. B. Treiman, R. Jackiw and D. J. Gross in: {\it Lectures on Current
Algebra and its Applications}, (Princeton, New Jersey, 1972).

\bibitem{dm2} D. Bisello {\it et al.} (DM2 Collab.), Nucl. Phys. 
(Proc. Suppl.) {\bf B21} (1991) 11.

\bibitem{aleph} D. Buskulic {\it et al.} (ALEPH Collab.), Z. Phys. {\bf
C74} (1997) 263.

\bibitem{pdg} Particle Data Group, R. M. Barnett {\it et al.},
 Phys. Rev. {\bf D54} (Part I) (1996) 1.

\bibitem{abj} S. Adler, Phys. Rev. {\bf 177} (1969) 2426; J. S. Bell and
R. Jackiw, Nuovo Cimento {\bf 60} (1969) 37. 

\bibitem{correction} In ref.[1], the use of the full $F_{\gamma \pi^0} =
2 F_{\gamma \pi^0}^{I=1}$ form factor led to a factor of 4 overestimate of
all the given $V_Q \rightarrow \omega \pi^0$ branching ratios.

\bibitem{bl1} G. P. Lepage and S. J. Brodsky, Phys. Lett. {\bf B87} (1979)
359; Phys. Rev. {\bf D22} (1980) 2157.

\bibitem{cello} H.-J. Behrend {\it et al.} (CELLO Collab.), Z. Phys. {\bf
C49} (1991) 401.

\bibitem{cleo} V. Savinov {\it et al.} (CLEO Collab.), Proceedings of the
PHOTON 95 Workshop, Sheffield (1995), Eds. D. J. Miller {\it et al.},
World Scientific; J. Gronberg {\it et al.} (CLEO Collab.), Phys. Rev. {\bf
D57} (1998) 33.

\bibitem{wz} T. F. Walsh and P. Zerwas, Nucl. Phys. {\bf B41} (1972) 551.

\bibitem{bl2} G. P. Lepage and S. J. Brodsky, Phys. Rev. {\bf D24} (1981)
1808.

\bibitem{new} For discussions in the light of the recent data, see e.g. P.
Kroll and M. Raulfs, Phys. Lett. {\bf B387} (1996) 848; I. V. Musatov and
A. V. Radyushkin, Phys. Rev. {\bf D56} (1997) 2713.

\end{thebibliography}
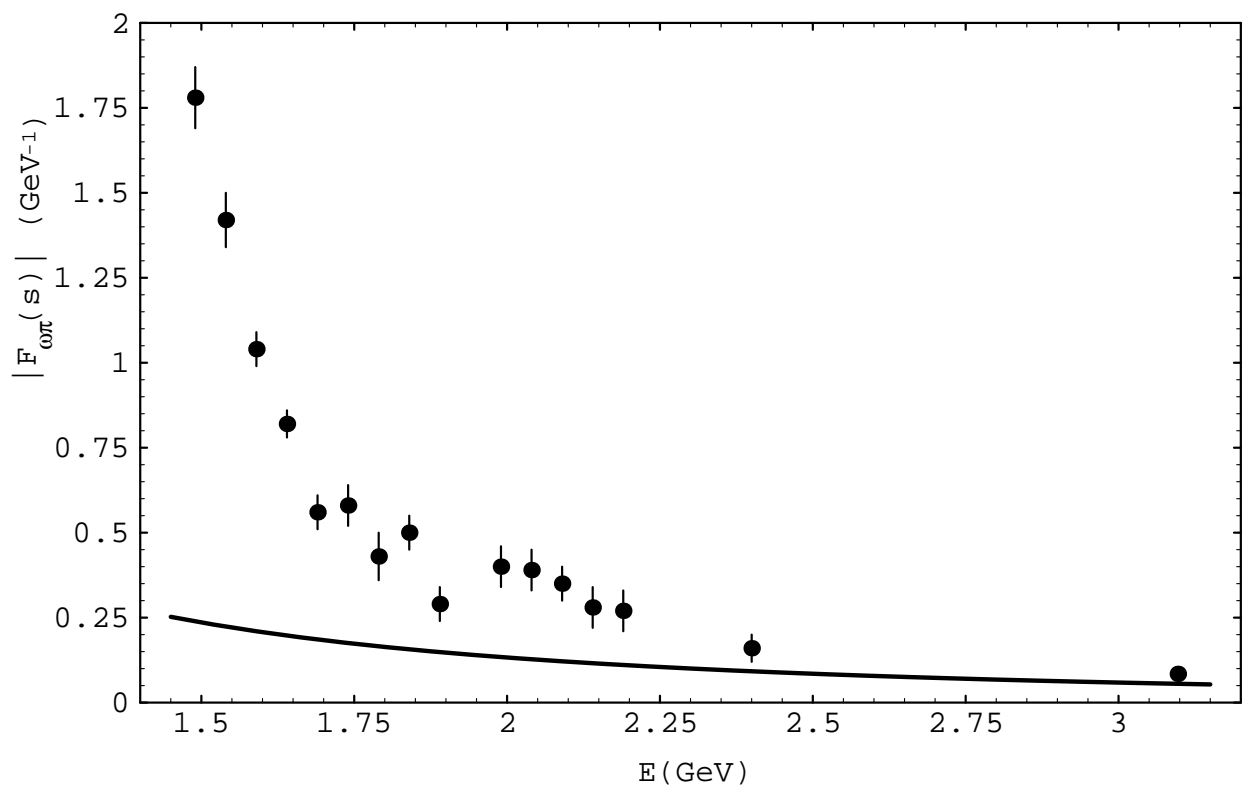
\end{document}